\documentclass[12pt,preprint]{aastex}

\usepackage{graphicx}
\usepackage{tipa}

\slugcomment{Revised ApJL, 2011 March 28}

\shorttitle{596 Scheila}
\shortauthors{Jewitt, et al.}

\begin{document}

\title{\emph{Hubble Space Telescope} Observations of\\
 Main Belt Comet (596) Scheila}

\author{David Jewitt$^{1,2,3}$, Harold Weaver$^4$, Max Mutchler$^5$, Stephen Larson$^6$, Jessica Agarwal$^7$  }
\affil{$1$ Dept. Earth and Space Sciences, UCLA, 595 Charles Young Drive East, Box 951567 Los Angeles, CA 90095-1567 \\
$2$ Institute for Geophysics and Planetary Physics, UCLA, 3845 Slichter Hall, 603 Charles Young Drive East, Los Angeles, CA 90065-1567 \\
$3$ Dept. Physics and Astronomy, UCLA, 430 Portola Plaza, 
        Los Angeles, CA 90095-1547 \\
$4$ The Johns Hopkins University Applied Physics Laboratory, 11100 Johns Hopkins Road, Laurel, Maryland 20723  \\
$5$ Space Telescope Science Institute, 3700 San Martin Drive, Baltimore, MD 21218 \\
$6$ Lunar and Planetary Laboratory, University of Arizona, 1629 E. University Blvd.
Tucson AZ 85721-0092 \\
$7$ Inst. for Physics and Astronomy, University of Potsdam, Karl-Liebknecht-Str. 24/25,
14476 PotsdamGermany \\
}
\email{contact: jewitt@ucla.edu}

\begin{abstract} 
We present \emph{Hubble Space Telescope Observations} of (596) Scheila during its recent dust outburst.  The nucleus remained point-like with absolute magnitude \mbox{$V(1,1,0)$ = 8.85$\pm$0.02} in our data, equal to the pre-outburst value, with no secondary fragments of diameter $\ge$100 m (for assumed albedos 0.04).   We find a coma having a peak scattering cross-section \mbox{$\sim$2.2$\times$10$^4$ km$^2$,} corresponding to a mass in micron-sized particles of $\sim$4$\times$10$^7$ kg.  The particles are deflected by solar radiation pressure on projected spatial scales $\sim$2$\times$10$^4$~km, in the sunward direction, and swept from the vicinity of the nucleus on timescales of weeks.  The coma fades by $\sim$30\% between observations on UT 2010 December~27 and 2011 January~04.  The observed mass loss is inconsistent with an origin either by rotational instability of the nucleus or by electrostatic ejection of regolith charged by sunlight.  Dust ejection could be caused by the sudden but unexplained exposure of buried ice. However, the data are most simply explained by the impact, at \mbox{$\sim$5 km s$^{-1}$,} of a previously unknown asteroid $\sim$35~m in diameter.   
\end{abstract}

\keywords{minor planets, asteroids: individual ((596) Scheila)); comets;}

\clearpage

\section{Introduction}

Main belt asteroid (596) Scheila (formerly 1906 UA and hereafter ``Scheila'') was discovered photographically  in 1906 by August Kopff.  It is a large body, with equivalent circular diameter $D$ = 113$\pm$2 km and a visual geometric  albedo 
\mbox{$p_v$ = 0.038$\pm$0.004}  (Tedesco et al.~2002).  The normalized (to 5500\AA) optical reflection spectrum of Scheila (Bus and Binzel 2004) has a gradient 
\mbox{$S'$ = 6.2$\pm$1\%/1000\AA\ }
(the uncertainty is our estimate of the systematic error; the statistical error is only 
\mbox{$\pm$0.04\%/1000\AA).}  The low albedo and the spectral slope together suggest that Scheila can be spectrally classified as a primitive P- or D- type (Dahlgren and Lagerkvist 1995).  Under the assumption of a bulk density 
\mbox{$\rho$ = 2000 kg m$^{-3}$} and a spherical shape, the approximate escape speed from Scheila is 
\mbox{$V_e$ = 60 m s$^{-1}$.}

\begin{figure}
\epsscale{1.00}
\begin{center}
\plotone{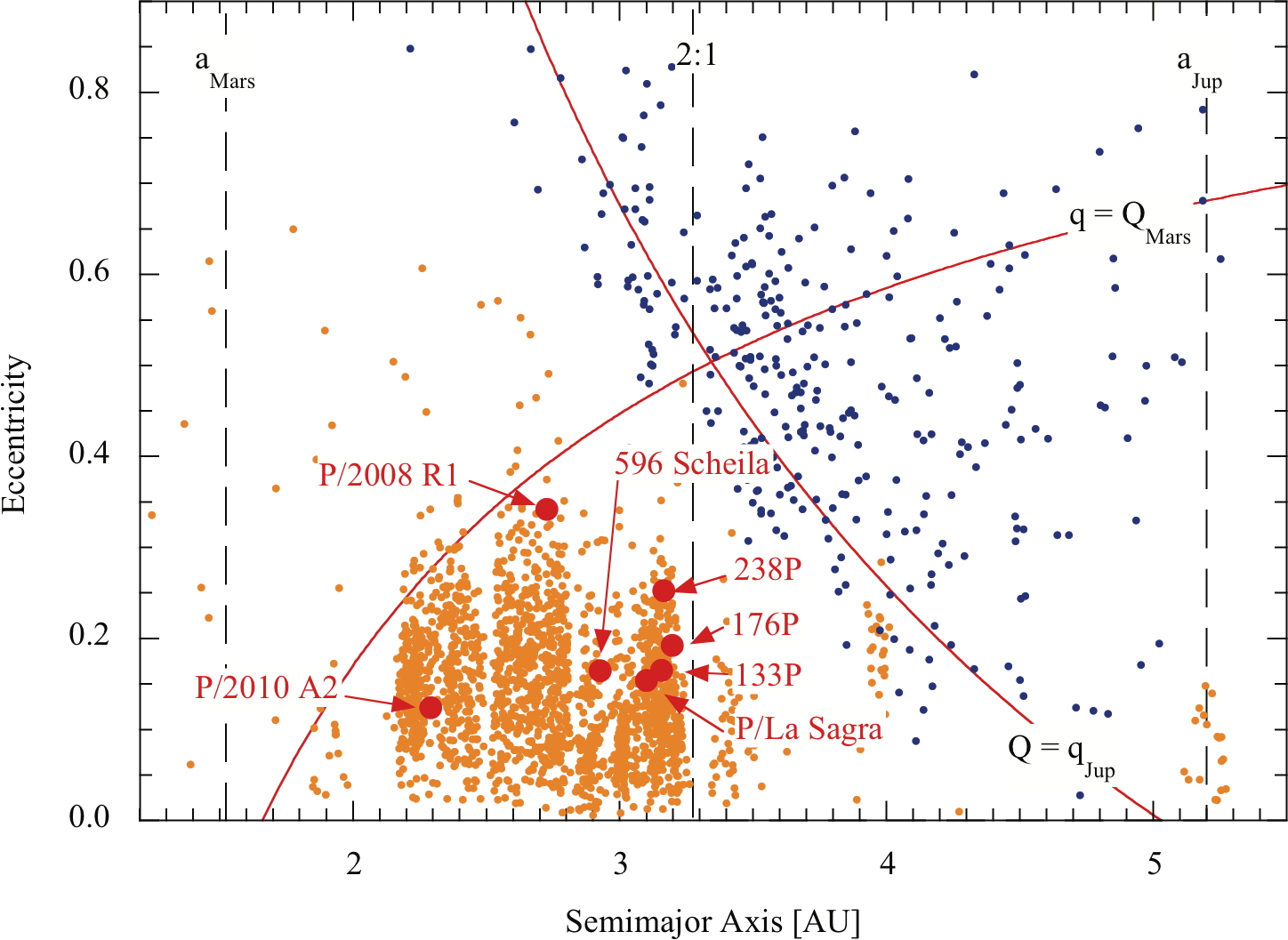}
\caption{Distribution of the known MBCs (red circles) in the semimajor axis vs. orbital eccentricity plane.  The corresponding distributions of asteroids (orange circles) and comets (blue circles) are shown for comparison.  Objects above the diagonal arcs cross either the aphelion distance of Mars or the perihelion distance of Jupiter, as marked.  The semimajor axes of the orbits of Mars and Jupiter are shown for reference, as is the location of the 2:1 mean-motion resonance with Jupiter.   \label{mbc_ae_plot} } 
\end{center} 
\end{figure}

Scheila's orbit lies in the outer asteroid belt (semimajor axis, $a$ = 2.926 AU, eccentricity, $e$ = 0.1644, and inclination, $i$ = 14.7$\degr$). The Tisserand parameter with respect to Jupiter, $T_J$ = 3.21, is typical of asteroids and lies far above the $T_J$ = 3 dynamical dividing line separating comets from asteroids (Kresak 1980).    However, Larson (2010) discovered a comet-like appearance in observations taken UT 2010 
\mbox{December 11.44-11.47} with the 0.68~meter Catalina Schmidt telescope.  Observations with the same telescope on UT 2010 \mbox{Dec 3.4} showed a slightly diffuse appearance and an integrated magnitude 
\mbox{$V$ = 13.2,} about 1.3 mag brighter than in the previous month, when the object appeared point-like (Alex Gibbs, reported in Larson 2010).  With a main-belt orbit \mbox{($T_J >$ 3)} and a comet-like morphology, Scheila satisfies the definition of a main-belt comet (MBC: Hsieh and Jewitt 2006). It is the seventh known example (and by far the largest) of this class (Figure \ref{mbc_ae_plot}). The MBCs appear to be of diverse origins, including examples likely to be driven by the sublimation of near surface ice (e.g., 133P/Elst-Pizarro; Hsieh and Jewitt 2006) and others likely to result from recent, probably collisional, disruption 
(e.g., P/2010~A2; Jewitt et al. 2010, Snodgrass et al. 2010).

Here we report initial observations with the \emph{Hubble Space Telescope (HST),} taken to 
examine Scheila at high angular resolution shortly after its outburst.

\section{Observations}
We secured two orbits of \emph{HST} Director's Discretionary Time on UT 2010 Dec. 27.9 and UT 2011 Jan 04.9 with the WFC3 camera~(Dressel et al. 2010).   On each orbit we took six exposures each of 4~s duration to image the near-nucleus environment and four exposures each of 390~s duration to examine the low surface brightness coma.  The short integrations employed the F621M filter (central wavelength $\lambda_c \sim$ 6210\AA~and FWHM [full width at half maximum] $\sim$ 640~\AA) to prevent saturation while maintaining the point-spread function of the image.  The long integrations used the F606W filter ($\lambda_c \sim$ 6000\AA~and FWHM  $\sim$ 2300~\AA) to maximize sensitivity to low surface brightness coma. At the time of observation the 0.04$\arcsec$ pixels corresponded to 66~km at the comet, so that the Nyquist sampled (2~pixel) resolution was 133 km.  A brief journal of observations is given in Table~\ref{journal}.

\begin{figure}
\epsscale{0.65}
\begin{center}
\plotone{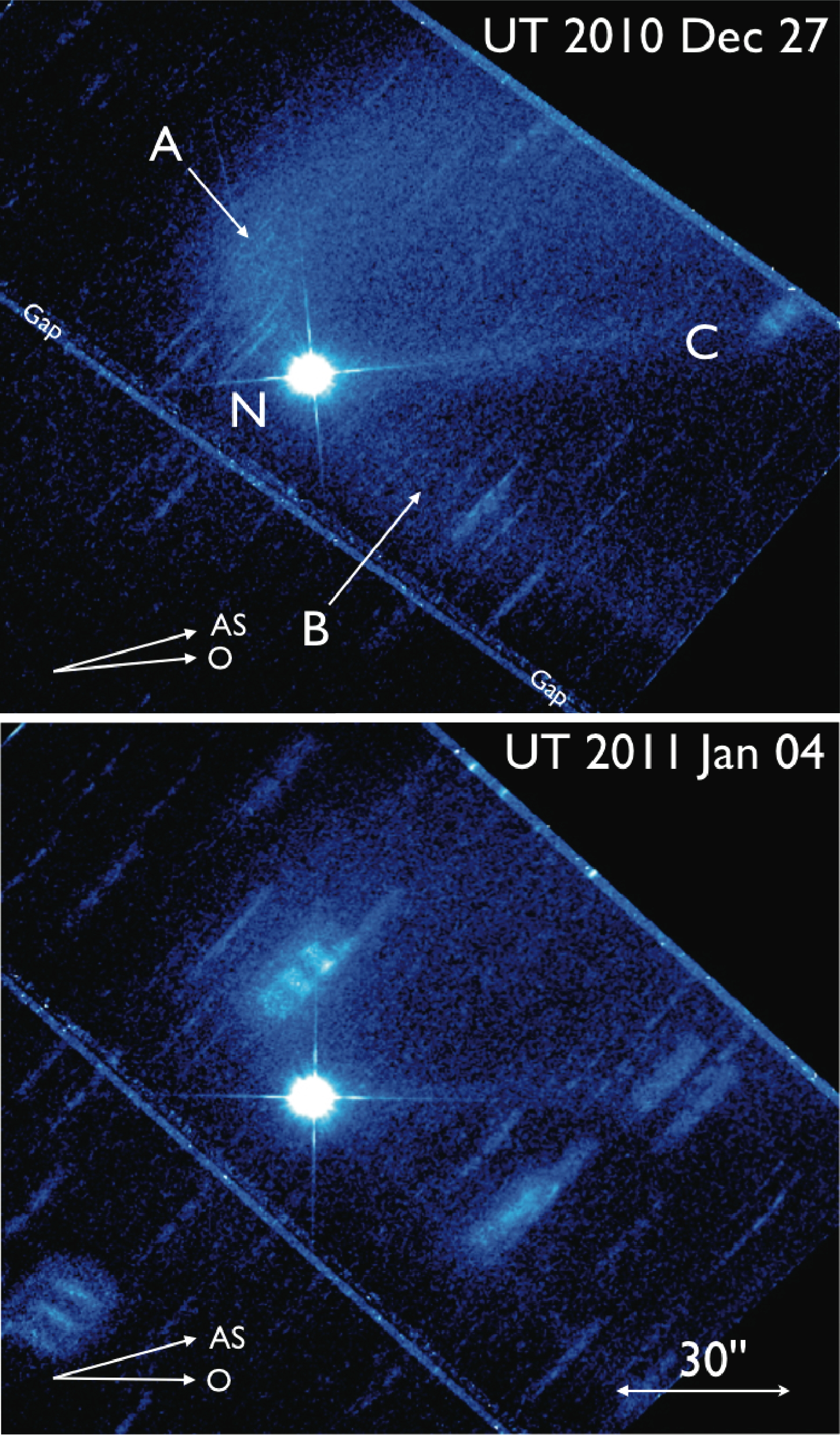}
\caption{Composite images from UT 2010 Dec~27 (top) and UT 2011 Jan~4 (bottom), each of 1560~s exposure.  The images have North to the top, East to the left and are shown with identical stretches. In the top panel, $N$ marks the nucleus, $A$ and
 $B$ are the north and south extensions of the small particle coma, and $C$ is the spike discussed in the text.  Marked arrows show (AS) the projected anti-solar direction and (O) the projected orbit.  The gap between chips in the detector is marked. \label{composite} } 
\end{center} 
\end{figure}

The images from each orbit were combined into composites having 24~s effective integration for the nucleus and 1560~s integration for the coma.  The nucleus ($N$ in Figure \ref{composite}) was found to be point-like, with FWHM = 2.2 pixels (0.09$\arcsec$). Photometry of the nucleus was obtained within a circular projected aperture of radius 0.4$\arcsec$, with sky subtraction from a concentric annulus with inner and outer radii 0.6$\arcsec$ to 0.8$\arcsec$, respectively 
(Table~\ref{photometry}).   The absolute magnitude, $V(1,1,0)$, (the magnitude if observed from  heliocentric and geocentric distances $R$ = $\Delta$ = 1 AU and at phase angle $\alpha$ = 0$\degr$) is computed from

\begin{equation}
V(1,1,0) = V - 5 \log_{10} \left(R \Delta \right) - f(\alpha).
\label{V110}
\end{equation}

\noindent  in which $f(\alpha)$ is the phase function.  For the latter we adopt the $H-G$ formalism of Bowell et al. (1989), with \mbox{$G$ = 0.076$\pm$0.060} as found from pre-outburst observations (Warner 2010).   As seen in Table~\ref{photometry}, the nucleus apparent magnitude brightened by $\sim$0.12~mag in the 8 day interval between the \emph{HST} observations, but this brightening is consistent with the changing observing geometry of the object, since $V(1,1,0)$ = 8.85$\pm$0.02 remains constant.  The derived absolute magnitudes from HST are consistent with the pre-outburst value, 8.84$\pm$0.04 (Warner 2010), showing that near-nucleus dust is insignificant.

On larger scales, the coma appeared diffuse and asymmetric, being brighter and more extended to the North than to the South of the nucleus on both visits ($A$ and $B$ in Figure \ref{composite}).  The sunward extension of the coma was $\sim$10$\arcsec$, corresponding to $\sim$16,000~km in the plane of the sky.  The overall form of the coma suggests that dust particles launched sunwards are being slowed by solar radiation pressure and so are concentrated near their turn-around points, giving rise to a bright-edged parabola (Figure \ref{composite}).  We connect $s$, the turn-around distance along the Sun-comet line to $u$, the initial sunward particle speed from

\begin{equation}
u^2 = 2 \beta g_{\odot} s
\end{equation}

\noindent where $\beta$ is the dimensionless radiation pressure factor and $g_{\odot}$ is the gravitational acceleration towards the Sun.   (Strictly, the data provide only a lower limit to $s$ because of the effects of projection (the phase angle was $\sim$13$\degr$, c.f. Table \ref{journal})). Substituting for $g_{\odot}$ we obtain

\begin{equation}
\beta = \frac{u^2 R^2}{2 G M_{\odot} s}
\label{beta}
\end{equation}

\noindent where $G$ is the gravitational constant, $R$ the heliocentric distance and $M_{\odot}$ the mass of the Sun.  The characteristic ejection velocity must be $u \geq V_e$, since slower particles should fall back to the nucleus under gravity.  Substituting $u \geq$ 60 m s$^{-1}$ in Equation \ref{beta} we obtain $\beta \geq$ 0.2.  The magnitude of $\beta$ is inversely related to particle size; \mbox{$\beta \geq$ 0.2} is compatible with dielectric particles having radii $\sim$0.1 to 1~$\mu$m (Bohren and Huffmann 1983).  We conclude that the diffuse coma of Scheila is populated by small dust grains just like those in other comets at 3 AU.  In this model, the characteristic travel time from the nucleus to the apex of motion is \mbox{$\tau$ = $2s/u$, or $\tau$ = 5$\times$10$^5$ s}
($\sim$1~week), while the residence time in the \emph{HST} field of view would be a month or more.   We note that the rapid fading of the coma on timescales of approximately a month strongly suggests that we are \textit{not} observing large ($>$10 $\mu$m) slow-moving grains in Scheila, in stark contrast to the situation for P/2010 A2 (Jewitt et al. 2010). 

To measure the brightness of the surrounding coma, we first digitally removed field star and galaxy image trails that were not already cancelled by the image combination process.   For this purpose we replaced afflicted pixels with the average pixel value measured in a surrounding region.  The coma brightness was measured using a projected circular aperture 64$\arcsec$ in radius, with sky subtraction from the median signal measured within a contiguous aperture having outer radius 80$\arcsec$.  The integrated magnitudes are brighter than the nucleus by about 
\mbox{$\Delta V$ = 1.3 mag}  on Dec 27 but only by  \mbox{$\Delta V$ = 1.0 mag} on 
Jan~04 (cf. Table~\ref{photometry}). 
This $\sim$30\% fading of the coma in the 8 days between measurements is broadly consistent with the timescale deduced above from radiation pressure sweeping, assuming the coma is not being continuously replenished.

We calculated the effective scattering cross-section of the coma  from
\mbox{$C_c = C_n (10^{0.4\Delta V} -1)$,} where 
\mbox{$C_n$ = $\pi r_n^2$ = 1.0$\times$10$^4$ km$^2$} is the geometric cross-section of the nucleus.  We further calculated the mass of dust particles in the coma from the scattering cross-section using

\begin{equation}
M_c = \rho \overline a C_c
\label{mass}
\end{equation}

\noindent where $\rho$ is the particle density, taken to be 
$\rho$ = 2000 kg m$^{-3}$, and $\overline a$ is the average particle radius in the coma.  We take 
$\overline a$ = 1 $\mu$m, corresponding to the upper limit estimated above from radiation pressure effects.  This size is consistent with the radiation pressure considerations discussed above and is also characteristic of optical observations of normal comets since much smaller particles are inefficient scatterers of optical photons while much larger particles are rare.   
The resulting dust cross-sections and masses are listed in Table~\ref{photometry}. The computed mass in micron-sized grains is strictly a lower limit to the total mass, since large particles may hold significant mass while presenting negligible scattering cross-section. Both $C_c$ and $M_c$ decreased between December~27 and January~4 by about 30\%, indicating that the escape of particles from the projected 64$\arcsec$ radius photometry aperture substantially exceeded the supply of fresh particles from the nucleus in this period.  

In addition to the diffuse coma, an approximately linear tail (the ``spike'') is evident (marked $C$ in Figure~\ref{composite}), with 
a position angle \mbox{277$\degr \pm 1\degr$} on Dec~27 and 
\mbox{273$\degr \pm 1\degr$} on Jan~04.  The position angles are slightly different from both the antisolar 
direction (position angles 276$\degr$ and 269$\degr$ on Dec~27 and Jan~04, respectively) and the projected orbit 
(position angle 286$\degr$ on both dates).  Therefore, it is not possible to interpret the spike as a simple synchrone, 
as was done for P/2010 A2.  Instead, it must consist of particles whose motion is determined by their initial velocity
as well as radiation pressure. 

We set upper limits to the allowable brightness of co-moving companions to Scheila by digitally adding scaled versions of the 
(unsaturated) nucleus.  At projected distances $\geq$6$\arcsec$ (10,000~km), 
any companion with $V \leq$ 28 could not escape detection.  Assuming that any companion has the same 
albedo as Scheila, a limit to the diameter of any secondary is set at \mbox{$d \leq$ 0.1 km.} 

\section{Discussion}
Two mechanisms considered previously as possible explanations for activity in main-belt comets are inoperable on 
Scheila, as a result of its large size and slow rotation.  First, mass loss through rotational instability is ruled out by the 
measured rotational period, \mbox{$P$ = 15.848 hr} (Warner 2006). 
The latter greatly exceeds the critical period at which the centripetal acceleration at the surface of a sphere equals 
the gravitational acceleration, assuming a bulk density \mbox{$\rho$ = 2000 kg m$^{-3}$.}
Second, the ejection of grains through electrostatic charging of the surface can be ruled out since the speeds 
generated electrostatically  \mbox{(v $\sim$1 m s$^{-1}$)} (Rennilson and Criswell 1974) are far smaller than the 
\mbox{$\sim$60 m s$^{-1}$} escape speed from the nucleus.  
Another mechanism must be responsible for the ejection of dust.

The simplest explanation is that Scheila ejected material after being struck by another, much smaller, asteroid.  This 
is the explanation proposed elsewhere for the inner-belt MBC P/2010~A2 (Jewitt et al. 2010).   In this interpretation, 
the estimated dust mass (Table~\ref{photometry}) gives a crude estimate of the impactor properties.  The velocity 
dispersion amongst main-belt asteroids is \mbox{$v_{i} \sim$5 km s$^{-1}$,}
about 10$^2$ times the escape velocity from 
Scheila.  At such a high speed, experiments show that a projectile can excavate 10$^3$ times its own mass from the 
target.  However, the bulk of the ejecta moves too slowly to escape.  Specifically, the mass of the ejecta leaving with 
\mbox{$v_e/v_i >$10$^{-2}$} is comparable to the impactor mass 
(Figure~4 of Housen and Holsapple 2011).  Therefore, the 
impact hypothesis requires that Scheila be struck by a body having mass 
\mbox{$M \sim$4.4$\times$10$^7$ kg} (Table~\ref{photometry}).
If also of density $\rho$ = 2000 kg m$^{-3}$, the impactor diameter would have been 
\mbox{$d_i \sim$ 35 m} and the kinetic energy of the impact \mbox{$\sim$5.5$\times$10$^{14}$ J}
(about 0.1~MTonnes TNT equivalent).
The resulting crater on Scheila would be of diameter \mbox{$\sim$ 400 m.}  

Another possibility is that the coma is produced by sublimation of surface ice, as in
\mbox{133P/Elst-Pizarro} (Hsieh and Jewitt 2006). 
Radio spectral line observations limit the production rate from Scheila to 
\mbox{log$_{10}$Q$_{OH} \leq$ 26.3} in the period December~14 - January~04, 
corresponding to mass production rates 
\mbox{$\leq$6 kg s$^{-1}$} in water (Howell and Lovell 2011).  However, these production limits are model-dependent, 
and cannot 
exclude the possibility that Scheila outgassed more strongly at earlier times, launching dust from the surface by the 
action of gas drag forces.

The energy balance equation for sublimating ice may be written

\begin{equation}
\frac{F_{\odot} \pi r_n^2 (1-A)}{R_{au}^2} = \chi \pi r_n^2 \left[ \epsilon \sigma T^4 + L(T) \frac{dm}{dt} \right]
\label{equilibrium}
\end{equation}

\noindent in which \mbox{$F_{\odot}$ = 1360 W m$^{-2}$} is the Solar Constant, 
$R_{au}$ is the heliocentric distance expressed in AU, $r_n$ (m) is the nucleus radius, 
$A$ is the Bond albedo, $\epsilon$ is the emissivity, $T$ (K) the effective temperature of the nucleus,
$L(T)$ (J kg$^{-1}$) the temperature-dependent latent heat of sublimation and 
\mbox{$dm/dt$ (kg m$^{-2}$ s$^{-1}$)} the rate of sublimation of the ice per unit area.  
Parameter $\chi$ in Equation (\ref{equilibrium}) is a proxy for the surface temperature variation over the surface of 
the nucleus, itself dependent on the nucleus shape, thermal properties, rotation period and spin axis direction 
relative to the Sun.  To solve Equation (\ref{equilibrium}) requires additional 
knowledge of $L(T)$ and $dm/dt$.  For this, we use the saturation vapor pressures and latent heat data for ice from 
Washburn (1926).  

Allowable values of $\chi$ lie in the range \mbox{1 $\leq \chi \leq$ 4.}
The high temperature limit, \mbox{$\chi$ = 1,} describes a flat plate oriented with its normal pointing towards the 
Sun, because then the absorbing and radiating areas are identical.  This would approximate, for instance, the 
heating of the sunward pole on a nucleus whose spin-vector points towards the Sun.  The cold limit, 
\mbox{$\chi$ = 4,} corresponds to an isothermal, spherical nucleus in which solar power is absorbed over an 
area \mbox{$\pi r_n^2$} but radiated from \mbox{$4\pi r_n^2$.}  
True cometary nuclei probably possess intermediate effective values of $\chi$.  For example, a 
spherical nucleus with the Sun on the polar axis corresponds to \mbox{$\chi$ = 2,} while a nucleus with the Sun in 
the equator and with rotation so rapid as to maintain temperature along lines of constant latitude has 
\mbox{$\chi$ = $\pi$,} and so on.  The point is that the rotational and thermophysical properties of a nucleus are 
usually unknown, so that the effective value of $\chi$, and thus the sublimation rate, 
cannot be accurately calculated.   

\begin{figure}
\epsscale{0.9}
\begin{center}
\plotone{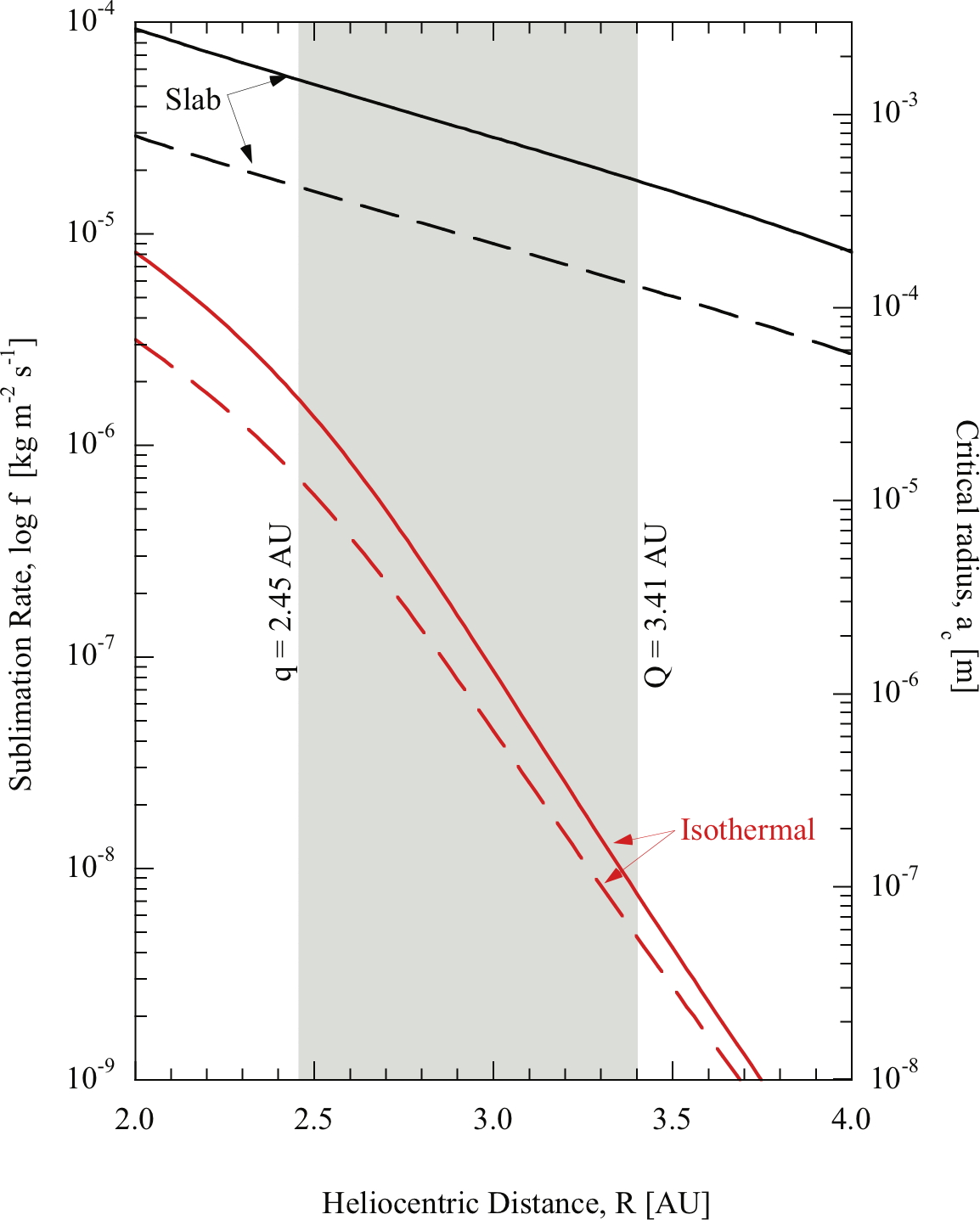}
\caption{Two-axis plot showing, on the left, the equilibrium water ice sublimation rate (solid lines) and, on the right, the maximum ejectable dust grain radius (dashed lines), both as functions of the heliocentric distance.  Red(black) curves correspond to the minimum(maximum) temperature extremes, as discussed in the text. The shaded region marks the range of heliocentric distances swept by Scheila each orbit.   \label{sublimation} } 
\end{center} 
\end{figure}

We used Equation~(\ref{equilibrium}) to calculate $dm/dt$ as a function of $R_{au}$ for spherical model comets with 
water ice exposed at the surface.  We used \mbox{$A$ = 0.03} and assumed 
\mbox{$\epsilon$ = 0.9.}  The limiting values \mbox{$\chi$ = 1} and \mbox{$\chi$ = 4} were assumed in order to 
bracket the range of likely behaviors to be expected from the real comets.  Results are shown in Figure~(\ref{sublimation}) for 
\mbox{2 $\leq R_{au} \leq$ 4 AU.}  Also shown in Figure~(\ref{sublimation}) is the critical radius, $a_c$, defined as 
the radius of the largest dust particle that can be ejected from Scheila against gravitational attraction to the nucleus, given sublimation at the rate computed from Equation (\ref{equilibrium}).  

Figure (\ref{sublimation}) shows that allowable temperature regimes on Scheila permit more than two orders of 
magnitude variation in the sublimation rate per unit area, 
from \mbox{$\sim$3$\times$10$^{-5}$ kg m$^{-2}$ s$^{-1}$} in the hot case 
to \mbox{$\sim$8$\times$10$^{-8}$ kg m$^{-2}$ s$^{-1}$} in the cold case, 
at \mbox{$R_{au}$ = 3 AU.}  If we assume that the dust/gas ratio of ejected material is of order unity, and that the 
dust ejection occurred over a period of less than 1 month \mbox{(2.5$\times$10$^6$ s),} then the area of exposed 
ice needed to supply the $M \sim$2$\times$10$^7$ kg coma mass is $A$ = 0.3 km$^2$ in the high temperature 
case rising to \mbox{$A$ = 100 km$^2$} in the low temperature case.   The surface area of Scheila is  
\mbox{$\sim$40,000 km$^2$,} so that even sublimation from the larger area corresponds to only 0.25\% of the 
surface.  

The critical dust radius for ejection at 3 AU varies from about 10$^{-4}$ m in the hot case to 
\mbox{6$\times$10$^{-6}$ m} in the cold case.  Thus, at this distance from the Sun, gas sublimated from exposed 
water ice is capable of launching dust particles above the escape speed no matter what the temperature regime on 
the surface.  Even at aphelion and with the lowest plausible surface temperatures, sublimated ice can eject grains 
with \mbox{$a_c \sim$ 0.1 $\mu$m,} the minimum size for efficient scattering of optical photons.  Therefore, both in 
terms of the area of exposed ice required, and in terms of the particle speeds induced by gas drag, ice sublimation 
is a plausible driver of the activity observed in Scheila at 3~AU, and potentially could supply observable coma at any 
position in the orbit.  

The ice hypothesis has two problems, however.  First, surface ice on Scheila is long-term unstable to sublimation and must be protected by a refractory mantle (Schorghofer~2008). What process exposed the ice in early December?  The most obvious possible trigger is an impact which, by the discussion above, would need to expose an area of ice 
\mbox{$A \geq$ 0.3 km$^2$} (corresponding to a circle 300~m in radius or larger).  But, as discussed above, an impact this large could produce an impulsive coma as observed even without the need for a sublimating ice source.  

Second, if ice sublimation is the driver of the activity, why does the coma fade on timescales of 
\mbox{$\tau \sim$ 8 days} \mbox{(7$\times$10$^5$ s)?} 
At \mbox{$dm/dt$ = 3$\times$10$^{-5}$ kg m$^{-2}$ s$^{-1}$,} ice would recede into the 
surface by a distance \mbox{$\delta$ = $\tau \rho^{-1} (dm/dt)$} which, by substitution gives 
\mbox{$\delta \sim$ 0.01 m (1 cm).}   This is smaller than the diurnal thermal skin depth 
\mbox{($d$ = 0.07 m [7 cm])} making it hard to see why the coma formation, once started, would so soon stop. 
For these two reasons, we conclude that the impact hypothesis provides the most simple and most likely explanation 
of the observed activity.  Scheila now stands, after P/2010 A2, as the second example of an impact-activated main-belt comet.


\section{Summary}
From images of (596) Scheila obtained using the \emph{Hubble Space Telescope} we find that:
\begin{enumerate}
\item The coma has effective peak cross-section \mbox{2.2$\times$10$^4$ km$^2$,} 
is shaped by radiation pressure, and fades on a timescale consistent with radiation pressure clearing. 

\item The coma dust particles have radiation pressure factors \mbox{$\beta >$0.2,} 
corresponding to micron-sized dielectric particles.  The mass in micron-sized dust particles is \mbox{4$\times$10$^7$ kg.}

\item The scattering properties of the nucleus appear unchanged by the ejection of dust.

\item No near-nucleus fragments or structures are apparent in the \emph{Hubble} data.

\item The measurements are consistent with dust ejection by impact into Scheila of a 
$\sim$35~m diameter projectile.


\end{enumerate}

\acknowledgements
Based on observations made with the NASA/ESA \emph{Hubble Space Telescope,}
with data obtained from the archive at the 
Space Telescope Science Institute (STScI). 
STScI is operated by the association of Universities for Research in Astronomy, Inc. 
under NASA contract NAS~5-26555. We thank the Director's Office at the STScI for granting us
Discretionary Time to make these observations and Alison Vick, Larry Petro, and other 
members of the STScI ground system team for their expert help in planning and scheduling these 
observations. Bin Yang and an anonymous referee supplied useful
comments on the manuscript.
DJ thanks Mike Hicks of JPL for supplying his Palomar observations.  Support is appreciated 
 from NASA's Planetary Astronomy program to DJ.

\clearpage

\begin{deluxetable}{llllcr}
\tablecaption{Journal of Observations
\label{journal}}
\tablewidth{0pt}
\tablehead{
\colhead{Instrument} &\colhead{UT\tablenotemark{a}}   & \colhead{$R$\tablenotemark{b}} & \colhead{$\Delta$\tablenotemark{c}}   & \colhead{$\alpha$\tablenotemark{d}} & \colhead{$\theta_i$\tablenotemark{e}} }
\startdata
WFC3/F606W/F621N & 2010 Dec 27.9 &              3.085  &    2.338   & 13.7 & -2.6 \\
WFC3/F606W/F621N & 2011 Jan 04.9 &                    3.073   &    2.254   & 11.9 & -3.5  \\

\enddata


\tablenotetext{a}{UT Date of the observation}
\tablenotetext{b}{Heliocentric distance in AU}
\tablenotetext{c}{Geocentric distance in AU}
\tablenotetext{d}{Phase (sun-object-Earth) angle in degrees}
\tablenotetext{e}{Out-of-plane angle in degrees}

\end{deluxetable}

\clearpage

\begin{deluxetable}{lllllcr}
\tablecaption{Photometry
\label{photometry}}
\tablewidth{0pt}
\tablehead{
\colhead{Date} &\colhead{$V_N$\tablenotemark{a}}   & \colhead{$V(1,1,0)$\tablenotemark{b}} & \colhead{$V$\tablenotemark{c}} & \colhead{$\Delta V$  \tablenotemark{d}}& \colhead{$C_c$\tablenotemark{e}}   & \colhead{$M$\tablenotemark{f}}  }
\startdata
2010 Dec 27.9 &           13.98$\pm$0.02  &  8.84$\pm$0.02&  12.63   & 1.26 & 2.2$\times$10$^4$ & 4.4$\times$10$^7$ \\
2011 Jan 04.9 &            13.86$\pm$0.02     &  8.86$\pm$0.02&  12.86   & 1.00 & 1.5$\times$10$^4$ & 3.0$\times$10$^7$  \\

\enddata


\tablenotetext{a}{Nucleus $V$-band magnitude measured within a 0.4$\arcsec$ radius aperture.  The apparent V-band magnitude was computed from the observed count rate ``C" in electrons s$^{-1}$) using $V = -2.5 \log C + Z$, where $Z$ = 24.45 for the F621M filter and 25.99 for the F606W filter (Kalirai et al. 2009).}
\tablenotetext{b}{Absolute $V$ magnitude of the nucleus (Equation \ref{V110})}
\tablenotetext{c}{Total $V$-band magnitude measured within a 64$\arcsec$ radius aperture}
\tablenotetext{d}{$\Delta V = V_N - V$}
\tablenotetext{e}{Effective scattering cross-section in km$^2$}
\tablenotetext{f}{Effective dust mass, in kg}

\end{deluxetable}

\clearpage

\clearpage

\clearpage

\clearpage


\clearpage



\begin{thebibliography}{}

\bibitem[Bohren 
\& Huffman(1983)]{1983asls.book.....B} Bohren, C.~F., \& Huffman, D.~R.\ 1983, New York: Wiley, 1983,  

\bibitem[Bowell et al.(1989)]{1989aste.conf..524B} Bowell, E., Hapke, B., 
Domingue, D., Lumme, K., Peltoniemi, J., 
\& Harris, A.~W.\ 1989, Asteroids II, 524 


\bibitem[Bus 
\& Binzel(2004)]{2004PDSS....1.1116B} Bus, S., \& Binzel, R.~P.\ 2004, NASA Planetary Data System, 1, 1116 

\bibitem[Dahlgren 
\& Lagerkvist(1995)]{1995A&A...302..907D} Dahlgren, M., \& Lagerkvist, C.-I.\ 1995, \aap, 302, 907 

\bibitem[Dressel et al (2010)]{D} Dressel, L., 2010. ÒWide Field Camera 3 Instrument Handbook, Version 3.0Ó (Baltimore: STScI) 


\bibitem[Housen 
\& Holsapple(2011)]{2011Icar..211..856H} Housen, K.~R., \& Holsapple, K.~A.\ 2011, Icarus, 211, 856 

\bibitem[Howell \& Lovell(2011)]{2011IAUC.9191....2H} Howell, E.~S., \& Lovell, A.~J.\ 2011, \iaucirc, 9191, 2 

\bibitem[Hsieh 
\& Jewitt(2006)]{2006Sci...312..561H} Hsieh, H.~H., \& Jewitt, D.\ 2006, Science, 312, 561 



\bibitem[Jewitt et al.(2010)]{2010Natur.467..817J} Jewitt, D., Weaver, H., 
Agarwal, J., Mutchler, M., \& Drahus, M.\ 2010, \nat, 467, 817 


\bibitem[Kalirai et al.(2009)]{2009STSCI} Kalirai, J. et al. 2009. WFC3 instrument Science Report 2009-31 (Baltimore: STScI)


\bibitem[Kresak(1980)]{1980M&P....22...83K} Kresak, L.\ 1980, Moon and Planets, 22, 83 

\bibitem[Larson(2010)]{2010IAUC.9188....1L} Larson, S.~M.\ 2010, \iaucirc, 
9188, 1 

\bibitem[Porco et al.(2005)]{2005Sci...307.1237P} Porco, C.~C., et al.\ 
2005, Science, 307, 1237

\bibitem[Rennilson 
\& Criswell(1974)]{1974Moon...10..121R} Rennilson, J.~J., \& Criswell, D.~R.\ 1974, Moon, 10, 121 


\bibitem[Schorghofer(2008)]{2008ApJ...682..697S} Schorghofer, N.\ 2008, 
\apj, 682, 697 

\bibitem[Snodgrass et al.(2010)]{2010Natur.467..814S} Snodgrass, C., et 
al.\ 2010, \nat, 467, 814 


\bibitem[Tedesco et al.(2002)]{2002AJ....123.1056T} Tedesco, E.~F., Noah, 
P.~V., Noah, M., \& Price, S.~D.\ 2002, \aj, 123, 1056 


\bibitem[Warner(2006)]{2006MPBu...33...58W} Warner, B.~D.\ 2006, Minor 
Planet Bulletin, 33, 58 

\bibitem[Warner(2010)]{2010cbet} Warner, B.~D.\ 2010, Central Bureau for Astronomical Telegrams, No. 2590 (December 15) 

\bibitem[Washburn(1926)]{1926washburn} Washburn, E.\ 1926, International Critical Tables of Numerical data, Physics, Chemistry and Technology, Vol. 3 (New York: McGraw-Hill).


\end{thebibliography}
\end{document}